Analysis of the $E_1$ and $E_1+\Delta_1$ optical transitions in (Ga,Mn)As epitaxial layers

L. Gluba[1]*, O. Yastrubchak[1], G. Sęk[2], W. Rudno-Rudziński[2], J. Sadowski[3,4], M. Kulik[1,5], W. Rzodkiewicz[6], M. Rawski[7], T. Andrearczyk[3], T. Wosinski[3] and J. Żuk[1]

1. Institute of Physics, Maria Curie-Skłodowska University, pl. M. Curie-Skłodowskiej 1, 20-031 Lublin, Poland
2. Institute of Physics, Wroclaw University of Technology, Wybrzeże Wyspiańskiego 27, 50-370 Wroclaw, Poland
3. Institute of Physics, Polish Academy of Sciences, Al. Lotników 32/46, 02-668 Warsaw, Poland
4. MAX-IV Laboratory, Lund University, P.O. Box 118, SE-22100 Lund, Sweden
5. Frank Laboratory of Neutron Physics, Joint Institute for Nuclear Research, 141980 Dubna, Russia
6. Institute of Electron Technology, Al. Lotników 32/46, 02-668 Warsaw, Poland
7. Faculty of Chemistry, Maria Curie-Skłodowska University, pl. M. Curie-Skłodowskiej 3, 20-031 Lublin, Poland

*e-mail: gluba@hektor.umcs.lublin.pl

Abstract

The diluted (Ga,Mn)As became a model ferromagnetic semiconductor, however there is still a disagreement on the source of its magnetism. The divergences arise from the results indicating that the holes mediated ferromagnetism reside in the valence band or the impurity band. Full understanding of character of the Mn states in GaAs can bring the increase of (Ga,Mn)As Curie temperature. In this paper we verify the ellipsometric results and compare with more precise photoreflectance method which gives a new insight into the interactions of Mn impurity states with GaAs valence band. Indeed, $E_1$ and $E_1+\Delta_1$ inter-band transition energies for highly doped and annealed (Ga,Mn)As epitaxial layers have not confirmed the interaction between detached Mn impurity band and the valence band. Thus, the description with merged Mn states and GaAs valence band is in agreement with our results. Our findings are supported by the high resolution transmission microscopy and magnetization measurements.

The diluted magnetic semiconductor (Ga,Mn)As, combining semiconducting properties with magnetism, belongs to the most widely investigated materials nowadays. Magnetic properties of (Ga,Mn)As give us an opportunity to test ideas for future semiconductor spintronics, integrating modern microelectronics with spin physics. The main advantage of diluted magnetic semiconductors is the fact that their magnetic properties can be successfully controlled by electric field [1] [2]. Despite great advances in epitaxial growth and post-growth low-temperature annealing of (Ga,Mn)As layers their highest Curie temperature $T_C$ = 185 K, as measured by superconducting quantum interference device (SQUID) magnetometry [3], is so far much below room temperature, desirable for practical applications.

The nature of carrier-mediated ferromagnetic ordering of $Mn^{2+}$ ions in (Ga,Mn)As has already been successfully explained over a decade ago within the frames of the *p-d* Zener model for ferromagnetic semiconductors [4]. However, theoretical understanding of the band structure and the underlying ferromagnetic interactions are still under debate [5] [6]. Two alternative models of the band structure in (Ga,Mn)As are predominant now. The Zener model, assuming *p-d* exchange interaction between the valence-band holes and localized Mn-3*d* electrons, predicts mixing of the impurity band arising from the Mn cations with the GaAs valence band providing for the valence-band origin of mobile holes mediating ferromagnetic ordering and the Fermi level position determined by the concentration of valence-band holes [4] [7]. On the other hand, the impurity-band model predicts that the Fermi level resides inside the Mn-related impurity band detached from the GaAs valence band [8] [9]. In this model ferromagnetism spreads within this impurity band by a hopping effect without affecting the valence-band holes. Not only theoretical models, but also experimental results, revealing different findings for similar (Ga,Mn)As layers, are controversial.

Although (Ga,Mn)As is a semiconductor with rather mediocre optical properties, a number of magneto-optical and optical methods have been employed to study its band structure. Among them, results of Faraday rotation and Kerr effect spectroscopy [10], magnetic circular dichroism [11] [12], spectroscopic ellipsometry (SE) [13] [14] and modulation photoreflectance (PR) spectroscopy [15] [16] [17] [18] investigations have been recently reported. The published results of SE measurements in the vicinity of $E_1$ and $E_1+\Delta_1$ band-gap optical transitions in (Ga,Mn)As layers provide with some contradictory conclusions. In particular, Burch et al. [13] reported on the ellipsometric investigations for (Ga,Mn)As layers with a wide range of Mn content of 0 to 6.6%. The authors suggested that the peak corresponding to the $E_1$ interband transition, which was present in the complex part of dielectric function spectra, was blue-shifted with increasing Mn content in respect to that of the GaAs substrate. Moreover, both the $E_1$ and $E_1+\Delta_1$ spectral features broaden under Mn doping and they merge just at a small (1%) Mn content. The authors claim that these results are in agreement with the impurity-band model and the blue-shift of $E_1$ transition results from *p-d* hybridization of the Mn induced impurity band and the GaAs valence band.

Another paper, showing the results of SE measurements for (Ga,Mn)As layers with an even wider range of Mn content of 0 to 9%, was published almost at the same time by Kang et al. [14]. Contrary to the previous results, neither a blue-shift of the $E_1$ transition energy nor a merging the $E_1$ and $E_1+\Delta_1$ peaks, with increasing Mn content, were revealed in the second paper. Kang et al. [14] found a small red-shift of the $E_1$ transition energy and a constant $E_1+\Delta_1$ band-gap energy with increasing Mn content. Those authors interpreted their results in terms of the *p-d* Zener model. In the two above-mentioned papers the investigated (Ga,Mn)As layers displayed similar magnetic properties but different SE responses and their contradictory interpretations.

Modulation PR spectroscopy has been successfully used in recent investigations of (Ga,Mn)As layers by some of the present authors [16] [17] [18]. In our papers the evolution of the fundamental optical band gap $E_0$, at the Γ point of the Brillouin zone, in (Ga,Mn)As layers with increasing Mn content, in the wide range from 0 to 6%, was studied. The results demonstrating a red-shift of the $E_0$ energy with increasing Mn content in the as-grown (Ga,Mn)As layers were interpreted as a result of merging the Mn-related impurity band with the GaAs valence band [16] [17]. Moreover, an increase in the band-gap transition energy caused by post-growth annealing treatment of the (Ga,Mn)As layers was interpreted as a result of annealing-induced enhancement of the free-hole concentration and the Fermi level location within the valence band [18], in accordance with the Zener model. The results of our measurements are consistent with the recent ab-initio calculations of the (Ga,Mn)As band-gap energy by Pela et al. [19], the experimental results of hard X-ray angle-resolved photoemission spectroscopy [20], and the very recent results of resonant photoemission spectroscopy supported by dynamical mean field theory [21]. On the other hand, PR investigations carried out by Alberi et al. [15] for (Ga,Mn,Be)As layers, with the Mn content in the range from 2.5% to 3.8%, were interpreted by assuming formation of the Mn-related impurity band due to the valence band anti-crossing phenomenon. However, those measurements were made for a rather narrow range of Mn content and, moreover, beryllium as an acceptor dopant in (Ga,Mn)As produces a large amount of Mn interstitials, which consequently destroy ferromagnetism in those layers [22].

In the present paper, in order to elucidate the controversial SE results, we have employed both the SE and modulation PR spectroscopy in the photon energy range of the $E_1$ and $E_1+\Delta_1$ critical points in (Ga,Mn)As. The $E_1$ and $E_1+\Delta_1$ band-gap transitions take place at the $\boldsymbol{k}$ wave vector near the L point, along the Λ directions (corresponding to the ⟨111⟩ crystallographic directions), of the Brillouin zone, where the conduction and valence bands are nearly parallel. The PR spectroscopy has an advantage over the SE as it is not affected by such factors as the presence of native oxide layers and their roughness, depth inhomogeneities, etc., influencing the ellipsometric measurements.

**Results**

We investigated six (Ga,Mn)As layers, with the Mn content from 1% to 6% and 20 nm − 300 nm thickness, grown by a low-temperature molecular-beam epitaxy (LT-MBE) technique at a temperature of 230°C on semi-insulating (001) GaAs substrates. In addition, we investigated, as a reference LT-GaAs layer, an undoped GaAs layer of the thickness of 230 nm, grown on GaAs by LT-MBE under the same conditions as the (Ga,Mn)As layers. Both the Mn compositions and the layer thicknesses were estimated during growth by the reflection high-energy electron diffraction (RHEED) oscillatory amplitude analysis.

As a result of relatively low growth temperature, LT-GaAs layers contain a large amount of about 1% of excess arsenic, mainly in a form of arsenic antisites, $As_{Ga}$, with a typical concentration of $1\times10^{20}$ cm$^{-3}$ [23]. In as-grown (Ga,Mn)As layers with a high Mn content, in addition to $Mn_{Ga}$ substitutional atoms, acting as acceptors in GaAs and supplying mobile holes, Mn interstitials are incorporated in a high concentration in the crystal lattice. The Mn interstitials, similarly as $As_{Ga}$ defects, act as double donors causing an effective compensation of mobile holes in the layers and a decrease in their Curie temperature. Post-growth annealing of the layers at temperatures below the LT-MBE growth temperature, resulting mainly in out-diffusion of the Mn interstitials, significantly increases their $T_C$ [24]. In order to improve structural quality and optical response of the investigated (Ga,Mn)As layers and to increase their $T_C$, two of the layers, of 6% Mn content and 20 nm and 100 nm thicknesses, have been subjected to the post-growth annealing treatment for 6 h and 60 h, respectively, performed in air at the temperature of 180ºC.

Properties of the investigated layers have been examined using several experimental techniques. The Raman spectroscopy and high-resolution x-ray diffraction results for the 230-nm-thick layers have been reported in our previous paper [16] showing a good structural quality of the layers, which were grown pseudomorphically on GaAs substrate under compressive misfit strain. Examination of the magnetic properties with SQUID magnetometry revealed ferromagnetic ordering at low temperatures in all the (Ga,Mn)As layers. The unannealed layers displayed the $T_C$ values in the range from 25 K to 60 K for the layers of 1% and 6% Mn content, respectively. The (Ga,Mn)As layers of 6% Mn content subjected to the post-growth annealing displayed distinctly higher $T_C$ of 73 K and 145 K for the 100-nm- and 20-nm-thick layers, respectively. High-resolution transmission electron microscope (HR-TEM) FEI TITAN was used to characterize the subsurface layer crystalline structure. Moreover, we have performed the nuclear reaction analysis of $O^{16}(\alpha,\alpha)O^{16}$ with the resonant energy 3.045 MeV [25] in order to measure the thicknesses of native oxides on the investigated layers, which remained below 0.5 nm in all the layers.

The SE measurements have been made at room temperature using the J.A. Woollam VASE (Variable Angle Spectroscopic Ellipsometer) configured with the rotating analyzer. The measured signal in the spectroscopic ellipsometry method comes from the near-surface part of the investigated layer, depending on penetration depth of the applied polarized light. SE is a convenient method for characterization of many materials, as it gives an opportunity to investigate a wide range of materials, no matter of surface conditions and the layer structure. From SE measurements the spectral dependences of the ellipsometric angles $\Psi(\lambda)$ and $\Delta(\lambda)$ are obtained, which enable to calculate the effective complex dielectric function $\varepsilon_1 + i\varepsilon_2$. Despite the fact that the principles of SE measurements are rather simple, the analysis of experimental results is usually complicated, especially in the case of new materials or inhomogeneous layers. In order to obtain the ($\varepsilon_1$, $\varepsilon_2$) spectra of a near-surface layer of a sample one has to create a sample structure model of the investigated layered material [26], containing such parameters as thicknesses, compositions and dielectric functions of the substrate and neighboring layers.

Modulation PR spectroscopy measurements were performed at room temperature using a set-up equipped with the excitation source of a Coherent 630 nm laser line, as a pump beam, and 250 W halogen lamp coupled to a monochromator, as a probe beam. The PR signal was collected by a Si photodiode with a quartz window. The measurement system was arranged in the so-called bright configuration with the light beam defocused on the investigated sample [27]. Photoreflectance spectroscopy is a modulation technique for probing optical transitions with a high accuracy even at room temperature. Contrary to spectroscopic ellipsometry, the PR spectroscopy requires a good crystalline quality of the investigated material in order to create electron – hole pairs with a mean free path large enough to modulate the dielectric function of the material. Earlier PR studies of LT-GaAs layers in the $E_1$ and $E_1+\Delta_1$ spectral region were not successful because, as the authors claimed, the epitaxial growth of GaAs at temperatures below 350°C resulted in too many defects in the investigated layers [28].

The crystalline quality of the (Ga,Mn)As epitaxial layers was analyzed using the HR-TEM method. Cross-sectional TEM images of two of the layers are presented in Fig. 1. The HR-TEM results demonstrated that the crystal structure of the (Ga,Mn)As layers was slightly distorted and this distortion increased with increasing the Mn content in the as-grown layers. However, the post-growth annealing treatment, which resulted mainly in out-diffusion of the Mn interstitials, distinctly improved the crystalline structure of the layers, as shown in Fig. 1b.

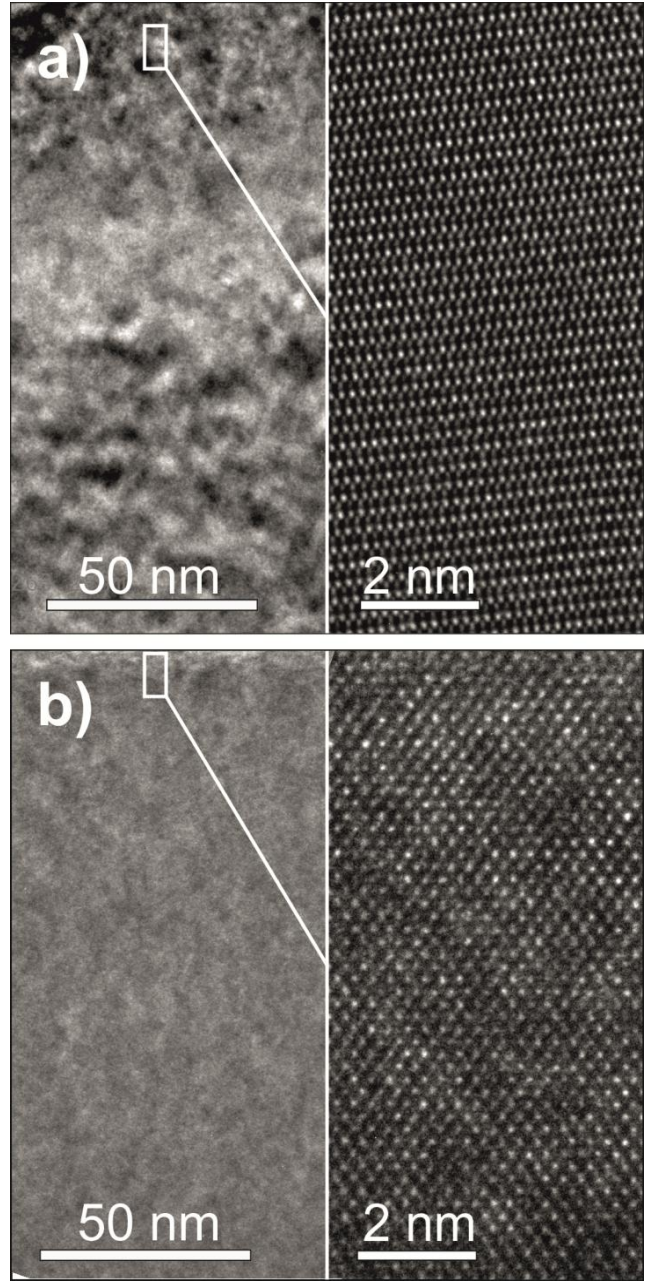

**Figure 1.** TEM cross-sectional images of the 300-nm-thick $Ga_{0.96}Mn_{0.04}As$ as-grown layer (a) and the 100-nm-thick $Ga_{0.94}Mn_{0.06}As$ layer subjected to post-growth annealing (b). High-resolution TEM images of the near-surface parts of the layers (marked with rectangles) are presented on the right-hand side of the figure.

Our SE results in the spectral range of the $E_1$ and $E_1+\Delta_1$ critical points for the 230-nm- and 300-nm-thick (Ga,Mn)As layers with the Mn content in the range from 0 to 6% are presented in Fig. 2. For all the layers the imaginary part of dielectric function spectra was evaluated using the VASE data collected at four incidence angles (65°, 70°, 75°, and 80°) for better computational accuracy of the dielectric function. Taking only optical absorption coefficient into account, the penetration depth of light can be defined as the reciprocal of absorption coefficient and can be expressed as:

$$d = \lambda\cos\theta/4\pi k, \qquad (1)$$

where $\lambda$, $\theta$ and $k$ are the wavelength of light, the incidence angle, and the extinction coefficient, respectively. In the photon energy range of the $E_1$ and $E_1 + \Delta_1$ critical points in GaAs the penetration depth value for $\theta = 75°$ is about 5 nm, which means that only top parts of the investigated layers were analyzed in our SE measurements.

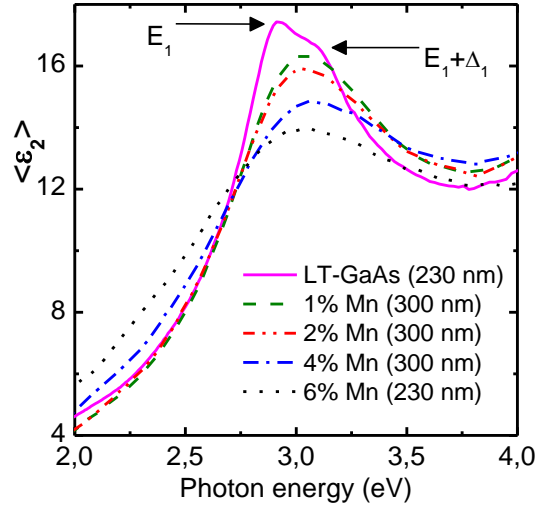

**Figure 2.** Imaginary part of dielectric function obtained from SE results for the LT-GaAs and (Ga,Mn)As layers. The Mn contents and layer thicknesses are written in the figure. Peaks corresponding to the $E_1$ and $E_1 + \Delta_1$ interband transitions in the LT-GaAs layer are marked with arrows.

SE spectra obtained for the (Ga,Mn)As/GaAs heterostructures are rather difficult for modeling by the VASE software. It is not easy to create a proper model for calculation of the complex dielectric function for such systems. Incorrect assumptions about surface quality can lead to wrong conclusions. In our model we have assumed, contrary to Burch et al. [13], that the thickness of native oxide on the (Ga,Mn)As layer surface was negligibly small (as it was estimated from the nuclear reaction analysis). The dielectric function has been obtained by means of the point-by-point calculation method using two phase (layer and substrate) model of the heterostructures. So-obtained spectra of the imaginary part of dielectric function are shown in Fig. 2. The peaks corresponding to the $E_1$ and $E_1 + \Delta_1$ interband transitions can be resolved for the LT-GaAs layer only. They broaden and merge into a single band for the (Ga,Mn)As layers at all Mn contents. This broad band is blue-shifted for the (Ga,Mn)As layers with increasing Mn content up to 4% (cross-sectional TEM images for that layer are shown in Fig. 1a). For the (Ga,Mn)As layer with 6% Mn content the band is red-shifted with respect to that of the layer with 4% Mn content.

Our attempts to obtain useful PR results in the photon energy range of the $E_1$ and $E_1 + \Delta_1$ optical transitions were not successful for the as-grown (Ga,Mn)As layers. The main reason for that was probably not sufficiently good crystalline quality of the layers, as previously suggested by Bernussi et al. [28] in their investigations of LT-MBE-grown GaAs layers. However, contrary to those previous investigations, we were able to obtain reasonable PR spectra for our reference LT-GaAs layer. Moreover, we succeeded in achieving the PR spectra of sufficient resolution for the (Ga,Mn)As layers subjected to post-growth annealing. As shown by our cross-sectional TEM results the annealing treatment resulted in significant improvement of the crystalline structure of (Ga,Mn)As layers.

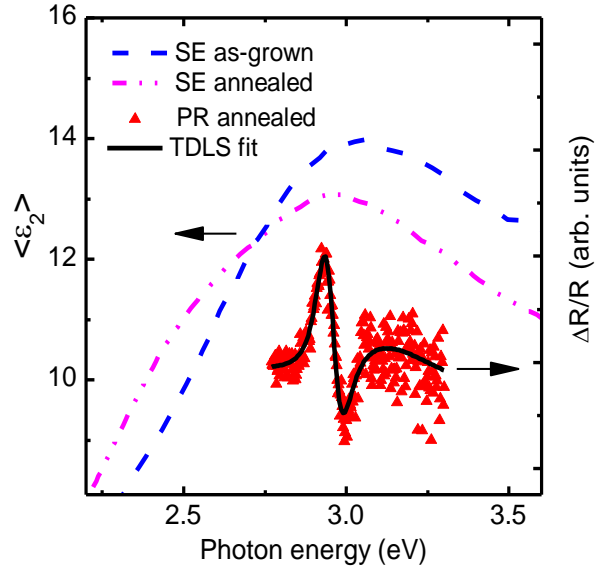

**Figure 3.** Imaginary part of dielectric function obtained from SE results for the as-grown and annealed 100-nm-thick $Ga_{0.94}Mn_{0.06}As$ layer (left scale) and the modulation PR spectrum for the annealed layer (right scale). Solid line represents the third derivative line shape (TDLS) fit to the experimental PR data.

Modulation PR spectrum obtained for the annealed 100-nm-thick (Ga,Mn)As layer with 6% Mn content is presented in Fig. 3. The SE results for the same layer, both as-grown and annealed, are also shown for comparison. Fig. 3 illustrates the difference in accuracy of determining the critical point energies from SE and PR results, and furthermore, it shows the effect of annealing the (Ga,Mn)As layer on its SE spectrum. As a result of the annealing treatment the resolution of ellipsometric spectra remains unimproved and much lower than that of PR spectrum for the same layer. The SE spectrum of the annealed (Ga,Mn)As layer displays a red-shift of its maximum with respect to that of the as-grown layer. However, it is not clear whether this effect results from a real shift of the merged $E_1$ and $E_1+\Delta_1$ band to lower energies or from some changes in the layer surface properties. As mentioned above, the SE technique, which relies on linear absorption mechanism, is very sensitive to the layer surface quality. On the other hand, the PR spectrum in the spectral range of $E_1$ and $E_1+\Delta_1$ critical points for the annealed (Ga,Mn)As layer, shown in Fig. 3, displays sufficient resolution to distinguish the $E_1$ spectral feature.

In Fig. 4 the PR spectra of the annealed (Ga,Mn)As layers of two thicknesses, of 20 and 100 nm, and the reference LT-GaAs layer are presented. The arrows indicate the $E_1$ and $E_1+\Delta_1$ transition energies obtained from the fitting procedure, to the experimental data, of the Lorentzian third derivative line shape (TDLS) formula for the relative changes in reflection coefficient given by Aspnes [29]:

$$\frac{\Delta R}{R} = Re[C\, e^{i\vartheta}(\hbar\omega - E_{CP} + i\gamma)^{-n}],  \quad (2)$$

where $C$ is an amplitude parameter, $\vartheta$ represents the phase, $\hbar\omega$ is the photon energy, $E_{CP}$ is the critical point energy, $\gamma$ is the broadening parameter and the index $n = 2.5$ for the three-dimensional band-to-band transition. Values of the $E_1$ and $E_1+\Delta_1$ critical point energies obtained from the TDLS fit to the experimental PR spectra, shown in Fig. 4, are listed in Table 1.

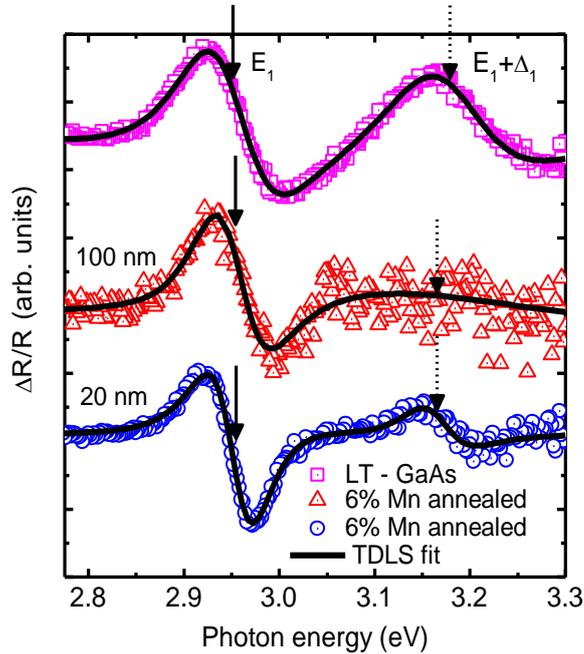

**Figure 4.** Modulation PR spectra for the LT-GaAs and annealed $Ga_{0.94}Mn_{0.06}As$ layers of 20 and 100 nm thickness. The spectra have been normalized to the same intensity and vertically offset for clarity. Solid lines represent the TDLS fits to the experimental data. Solid and dotted arrows indicate energetic positions of the $E_1$ and $E_1+\Delta_1$ transitions, respectively.

Despite the PR spectrum of the thinner, 20-nm-thick layer (lower spectrum in Fig. 4) is of much higher resolution than that of the thicker, 100-nm-thick one (middle spectrum in Fig. 4), the energy values of the two critical points obtained for the annealed (Ga,Mn)As layers are very similar. The higher resolution of the PR spectrum for the thinner layer likely results from more efficient outdiffusion of Mn interstitials during the annealing treatment of this layer with respect to the thicker one, which is also confirmed by significantly larger Curie temperature in the annealed 20-nm-thick (Ga,Mn)As layer, as mentioned above. When compared to the results obtained for the reference LT-GaAs layer, the energy values of the $E_1$ critical point are insignificantly blue-shifted for the annealed (Ga,Mn)As layers. On the other hand, the energies of the $E_1+\Delta_1$ critical point for the annealed (Ga,Mn)As layers are distinctly red-shifted with respect to that of the LT-GaAs layer.

Table 1. Values of the critical point energies for the LT-GaAs and annealed (Ga,Mn)As layers obtained from the TDLS fit to the PR spectra shown in Fig. 4.

| layer | $E_1$ (eV) | $E_1 +\Delta_1$ (eV) |
|---|---|---|
| LT-GaAs: 230 nm | 2.951 | 3.179 |
| (Ga,Mn)As: 6% Mn, 100 nm | 2.954 | 3.160 |
| (Ga,Mn)As: 6% Mn, 20 nm | 2.955 | 3.165 |

**Discussion**

Our PR spectroscopy results, showing, in contradiction to the SE results by Burch et al. [13], no significant blue-shift of the $E_1$ interband transition, cannot be explained within the impurity band model, which takes into account the *p-d* hybridization of the Mn induced impurity band and the GaAs valence band. As pointed out by Qi et al. [30], such hybridization should depend on the ***k*** wave vector and be very strong near the L point, leading, in consequence, to larger blue-shift of the $E_1$ interband transition than that of the $E_0$ transition. In this respect, our present results, together with our previous findings [16] [17] of a red-shift of the $E_0$ transition in (Ga,Mn)As layers with increasing Mn content, are consistent with the Zener model of ferromagnetism in (Ga,Mn)As. On the other hand, our present observation of well-resolved $E_1$ and $E_1+\Delta_1$ critical points in the PR spectra of (Ga,Mn)As layers confirms that the ***k*** wave vector remains a good quantum number in (Ga,Mn)As, at least in the annealed layers with Mn content up to 6%.

In summary, optical transitions corresponding to the $E_1$ and $E_1+\Delta_1$ critical points in (Ga,Mn)As epitaxial layers with the Mn content from 0 to 6% have been investigated by modulation photoreflectance spectroscopy and spectroscopic ellipsometry. Well-resolved PR spectra in the photon energy range of these transitions have been obtained for the first time for (Ga,Mn)As layers subjected to post-growth annealing treatment. The annealing resulted in an increase in the Curie temperature of the layers up to 145 K, as measured with the SQUID magnetometry, and an improvement of the layer crystalline quality, confirmed by the high-resolution TEM experiments. Mn incorporation into the GaAs crystal lattice resulted in merging the spectral features corresponding to the $E_1$ and $E_1 +\Delta_1$ critical points, observed in the imaginary part of dielectric function obtained from SE results for the reference LT-GaAs layer, into a single broad band and in a blue-shift of this band with increasing Mn content. However, this blue-shift was not observed in the SE results for the annealed (Ga,Mn)As layers displaying improved crystalline quality and higher Curie temperatures. Moreover, no blue-shift of the $E_1$ and $E_1 +\Delta_1$ transition energies in the annealed (Ga,Mn)As layers, with respect to those in LT-GaAs layer, was revealed from PR spectroscopy results, which were less affected by the layer surface quality than the SE results. Our PR results, showing no change of the $E_1$ and a small red-shift of the $E_1+\Delta_1$ transition energies in (Ga,Mn)As with respect to those in LT-GaAs, cannot be explained in terms of the impurity band model, which assumes persistence of the Mn-related impurity band detached from the valence band and an increase in the interband optical transition energies with increasing Mn content in (Ga,Mn)As.

[1] Chiba, D., Sawicki, M., Nishitani, Y., Nakatani, Y., Matsukura, F. & Ohno, H. Magnetization vector manipulation by electric fields. *Nature* **455** 515, (2008).

[2] Mark, S. et al. Scanning electron microscope image of a fully electrical read-write device made from a ferromagnetic semiconductor. *Phys. Rev. Lett.* **106** 057204 (2011).

[3] Wang, M. et al. Achieving high Curie temperature in (Ga,Mn)As. *Appl. Phys. Lett.* **93**, 132103 (2008).

[4] Dietl, T., Ohno, H., Matsukura, F., Cibert, J. & Ferrand, D. Zener model description of ferromagnetism in zinc-blende magnetic semiconductors. *Science* **287**, 1019–1022 (2000).

[5] Dietl, T. A ten-year perspective on dilute magnetic semiconductors and oxides. *Nat. Mater.* **9**, 965–974 (2010).

[6] Samarth, N. Ferromagnetic semiconductors: Battle of the bands. *Nat. Mat.*, **11** 360 (2012).

[7] Jungwirth, T., Mašek, J., Kučera, J. & MacDonald, A. H. Theory of ferromagnetic (III,Mn)V semiconductors. *Rev. Mod. Phys.* **78**, 809–864 (2006).

[8] Berciu, M., & Bhatt, R. N. Mean-field approach to disorder effects on ferromagnetism in (III,Mn)V at low carrier densities. *Phys. Rev. B* **69**, 045202 (2004).

[9] Burch, K. S., et al. Impurity band conduction in a high temperature ferromagnetic semiconductor. *Phys. Rev. Lett.* **97**, 087208 (2006).

[10] Acbas G., et al. Electronic Structure of Ferromagnetic Semiconductor Ga1-xMnxAs Probed by Subgap Magneto-optical Spectroscopy. *Phys. Rev. Lett.* **103**, 137201 (2009).

[11] Berciu, M., et al. Origin of magnetic circular dichroism in GaMnAs: giant Zeeman Splitting vs. spin dependent density of states. *Phys. Rev. Lett.* **102**, 247202 (2009).

[12] Jungwirth, T., et al. Systematic Study of Mn-Doping Trends in Optical Properties of (Ga,Mn)As Phys. Rev. Lett. **105**, 227201 (2010).

[13] Burch, K. S., Stephens, J., Kawakami, R. K., Awschalom, D. D. & Basov, D. N. Ellipsometric study of the electronic structure of $Ga_{1-x}Mn_xAs$ and low-temperature GaAs. *Phys. Rev. B* **70** 205208 (2004).

[14] Kang, T. D., Lee, G. S., Lee, H., Koh, D., & Park, Y. J. Optical Properties of $Ga_{1-x}Mn_xAs$ (0 ≤x≤ 0.09) Studied Using Spectroscopic Ellipsometry. *Korean Phys. Soc.* **46,** 482 (2005).

[15] Alberi, K., et al. Formation of Mn-derived impurity band in III-Mn-V alloys by valence band anticrossing. *Phys. Rev. B* **78,** 075201 (2008).

[16] Yastrubchak, O., et al. Photoreflectance study of the fundamental optical properties of (Ga,Mn)As epitaxial films. Phys. Rev. B **83,** 245201 (2011).

[17] Yastrubchak, O., et al. Electronic- and band-structure evolution in low-doped (Ga,Mn)As. *J. Appl. Phys.* **114**, 053710 (2013).

[18] Yastrubchak, O., et al. J. Appl. Phys. **115**, (Jan. 07, 2014, in print)

[19] Pelá, R. R., Marques, M., Ferreira, L. G., Furthmüller, J., and Teles, L. K., GaMnAs: Position of Mn-d levels and majority spin band gap predicted from GGA-1/2 calculations. *Appl. Phys. Lett.* **100,** 202408 (2012).

[20] Gray, A. X. et al. Bulk electronic structure of the dilute magnetic semiconductor $Ga_{1-x}Mn_xAs$ through hard X-ray angle-resolved photoemission. *Nat. Mater.* **11**, 957 (2012)

[21] Di Marco, I. et al. Electron correlations in $Mn_xGa_{1-x}As$ as seen by resonant electron spectroscopy and dynamical mean field theory. *Nat. Commun.* **4**, 2645 (2013).

[22] Yu, K. M. et al. Curie temperature limit in ferromagnetic $Ga_{1-x}Mn_xAs$. *Phys. Rev. B* **68**, 041308(R) (2003).

[23] Look, D. C., Walters, D. C., Manasreh, M. O., Sizelove, J. R., Stutz, C. E. & Evans K. R. Anomalous Hall-effect results in low-temperature molecular-beam-epitaxial GaAs: Hopping in a dense EL2-like band. *Phys. Rev. B* **42,** 3578 (1990).

[24] Edmonds, K. W. et al. Mn Interstitial Diffusion in (Ga,Mn)As. *Phys. Rev. Lett.* **92**, 037201 (2004).

[25] Cameron, J. R. Elastic Scattering of Alpha-Particles by Oxygen. *Phys. Rev.* **90**, 839 (1953).

[26] Dmitruk, N. L., Klopfleisch, M., Mayeva, O. I., Mamykin, S. V., Venger, E. F., & Yastrubchak, O. B. *Phys. Stat. Sol. A* **184**, 165 (2001)

[27] Misiewicz, J., Sitarek, P., Sęk, G. *Opto−Electron. Rev.* **8**, 1 (2000).

[28] Bernussi, A. A., Souza, C. F., Carvalho, W., Lubyshev, D. I., Rossi, J. C., & Basmaji, P. Optical and structural properties of low temperature GaAs layers grown by molecular beam epitaxy. *Braz. J. Phys.* **24**, 460 (1994).

[29] Aspnes, D. Third-derivative modulation spectroscopy with low-field electroreflectance. Surf. Sci. **37**, 418 (1973).

[30] Qi, J. et al. Mechanical and electronic properties of ferromagnetic $Ga_{1-x}Mn_xAs$ using ultrafast coherent acoustic phonons. *Phys. Rev. B* **81**, 115208 (2010).

Acknowledgements

OY acknowledges financial support from the Foundation for Polish Science under Grant POMOST/2010-2/12 sponsored by the European Regional Development Fund, National Cohesion Strategy: Innovative Economy. This work was also supported by the Polish National Science Centre under grant No. 2011/03/B/ST3/02457. The MBE project at MAX-IV Laboratory is supported by the Swedish Research Council (VR).